\documentstyle[12pt]{article}
\def\lc{\lambda _{\rm critical}}
\def\gc{g_{\rm critical}}
\begin{document}

\noindent\hfill  OHSTPY-HEP-TH-93-15\\
\begin{center}\begin{Large}\begin{bf}
The Light-Cone Field Theory Paradigm for Spontaneous Symmetry Breaking
\footnote{Talk given at Hadron Structure '93, Banska Strianvnica Slovakia,
September 5-10, 1993
based on Notes by Oliver Schnetz and Klaus Lucke, Institut f\"ur
theoretische Physik III, Staudtstr. 7, D-8520 Erlangen of lectures presented
to the Graduiertenkoleg Erlangen-Regensburg on May 26, 1993.  The work
presented
here was done in collaboration with C. M. Bender and B. van de Sande}\\
\end{bf}\end{Large}\end{center}
\vspace{.75cm}\begin{center}
Stephen Pinsky\\[10pt]
     \end{center}
      \vspace{0.1cm}
      \begin{center}
      \begin{it}
Department of Physics\\
The Ohio State University\\
174 West 18th Avenue\\
Columbus, Ohio  43210\\
       \end{it}
         \end{center}

\newpage

\tableofcontents

\section{Introduction}

In the first part of this lecture I will give an introduction to
light-cone field theory, focussing on the ``zero mode problem''. In the
second part I discuss $\phi^4$-theory in 1+1 dimensions. I will show how
the dynamics of the zero modes can give rise to spontaneous symmetry
breaking in spite of the trivial vacuum structure on the light-cone.

\section{ Review of Light-Cone Field Theory}

\subsection{Motivation}

One of the major outstanding problems in physics is how to calculate
obser\-vable processes in strongly interacting field theories like QCD and
electroweak theory. In particular we do not know, how to calculate from
first principles the hadronic spectrum, structure functions,
fragmentation functions, weak decay amplitudes and nuclear structure.
For now the two most promising attempts to tackle strongly interacting
field theories are lattice calculations and light-cone field theory.
This lecture is devoted to the light-cone approach\footnote{Many of the
topics discussed in this section are reviewed in [1].}.
In the 60's Fubini, Furlan and Weinberg [2] showed that in a Poincar\'e
invariant theory calculations may be simpler in an ``infinite momentum
frame'', i.e. in a frame moving with $v \rightarrow c$ relative to the
centre of mass. Weinberg showed that the singularities for $\gamma
=\frac{1}{\sqrt{1-\left(\frac{v}{c}\right)^2}} \rightarrow \infty$
cancel in the physical observables. The net effect (apart from a singular
scale factor) is to transform to light-cone coordinates
$$x^+:=\frac{1}{\sqrt{2}}(x^0+x^3) \qquad
  x^-:=\frac{1}{\sqrt{2}}(x^0-x^3) \qquad
  x_\perp:=(x^1,x^2) \quad \mbox{unaffected} $$
with $x^+$ regarded as the (light-cone) time and $x^-$, $x^1$ and $x^2$
regarded as spatial coordinates\footnote{I take the (+,-,-,-) metric and
$x^+$ (rather than $x_+=x^-$ as in [4]) as the time coordinate. }.
This interpretation is crucial as the
Hamiltonian formalism doesn't treat space and time in a symmetric way.

Note that strictly speaking the transformation to light-cone coordinates
is not a Lorentz transform. Yet it is generally believed, but not trivial
(and in fact not yet proven for the non-perturbative case), that the
quantum theory based on light-cone coordinates is equivalent to the
theory in ordinary coordinates.

\subsection{Quantization on the Light-Cone}

We are used to quantizing a theory by fixing commutation relations at
equal times.  But, as Dirac [3] pointed out in 1949, there are other
hypersurfaces in Minkowsky space that can be used for quantizing a
theory. Leutwyler and Stern [4] give a complete list and in
particular discuss light-cone quantization.
\medskip
The general procedure is described below:
\begin{enumerate}
\item
        Choose the hypersurface that you want to use for quantization
        (e.g. $x^+=0$)
\item
        Identify a complete set of independent dynamical variables
\item
        Set up the (hypersurface) commutation rules for the complete set
\item
        Identify the 10 generators of the Poincar\'e group in terms of
        your variables and check if they satisfy the Poincar\'e algebra
\end{enumerate}

\subsubsection{Zero Modes}

Let us now look at this procedure in more detail for the chosen
hypersurface being $x^+=0$. If massless particles are involved we can
not specify the boundary conditions for the classical dynamical
evolution as we can see in
the example of the massless Klein-Gordon equation in 1+1 dimensions
$$ \frac{\partial^2}{\partial t^2}\phi - \frac{\partial^2}{\partial
x^2}\phi =0  \quad.$$
Rewritten in light-cone coordinates
$$\frac{\partial}{\partial x^+} \frac{\partial}{\partial x^-} \phi =0$$
we can immediately read off the most general classical solution
$$\phi(x^+,x^-) = f(x^+)+g(x^-) $$
with arbitrary\footnote{W.l.g. we can assume $\int dx^- g(x^-)=0$ .}
$f$ and $g$.

If we try to fix the boundary conditions at $x^+=0$, we get almost no
information about $f(x^+)$ and in addition are not allowed to set up
conditions for the $x^+$-derivative for more than one point on the
surface. This is why mathematicians warn us to fix
the boundary conditions on characteristics (here the nullplanes are
characteristics of the Klein-Gordon equation).

I denote the (spatial) Fourier modes of $\phi$ as
$$ a(p^+,x^+):=\int dx^- e^{ip^+ x^-}\phi(x^+,x^-) .$$
The zero mode $a(p^+=0)$ corresponds\footnote{If $x^-$ is confined to an
interval, we can speak of discrete modes and
$a(p^+=0)=f*\mbox{interval length}$.} to $f$.
Hence about half of the massless degrees of freedom reside in the zero
mode. They correspond to propagation along the $x^-$-axis .

As we will see, the zero mode will give rise to spontaneous symmetry
breaking in the $\phi^4$ quantum theory.

\subsubsection{QCD}

To make you familiar with the procedure of finding  the dynamical
variables I briefly discuss QCD. First we write down the usual
Lagrangian (suppressing colour and flavour)
$$ {\cal L}=\bar\psi (i /\!\!\!\!\partial + g /\!\!\!\!A )\psi
-\frac{1}{2} tr F_{\mu \nu}F^{\mu \nu}    .$$
To identify the dynamical variables, we note that the projection
operators\footnote{Please don't confuse them with energy and momentum
that are denoted by the same symbols.}
$P^\pm := \frac{1}{2}\gamma^\mp \gamma^\pm = \frac{1}{2}(1 \pm \gamma^0
\gamma^3) $
fulfill
$$P^++P^-=1 \qquad \gamma^+ P^- =0 \qquad \gamma^- P^+ =0  \quad .$$
So $\psi^-:=P^- \psi $ doesn't appear in combination with an
$x^+$-derivative. Hence only $\psi^+:=P^+ \psi $ is a dynamical variable.
So in the gauge\footnote{I don't want to discuss the subtle questions
arising from the zero mode of $A_-$ in light-cone qantization [14].}
$A_- =0$ the only dynamical variables are $\psi^+$,$A_1$
and $A_2$. $\psi^-$ and $A_+$ have to be eliminated from the Lagrangian
(i.e. expressed in terms of the dynamical variables) via constraint
equations.

Now we demand canonical commutation relations for the dynamical
variables and their conjugate momenta. Actually the canonical
commutation relations in light-cone coordinates contain some unusual
features. One has to derive the commutators from the
Dirac-Bergmann
prescription or the Schwinger action principle. For this I refer to the
literature [7].

 Then we can construct the energy
momentum tensor $T^{\mu \nu}$ and the angular momentum desity $M^{\mu \nu
\lambda}$ in terms of our variables. Integration of the $\mu = +$
components over $\int dx^- dx_\perp$ yields the 10 generators of the
Poincar\'e group $P^\nu$ and $J^{\nu \lambda}$. For consistency one has
to check if the generators yield the correct (Poincar\'e algebra)
commutation relations.

\subsubsection{Kinematical Generators}

A major difference between quantization in ordinary coordinates and on
the light-cone is, that different generators of the Poincar\'e group become
simple.
In the equal-time formalism the generators of
(spatial) rotations ($J_{12}$, $J_{23}$, $J_{31}$) and
shifts ($P_1$, $P_2$, $P_3$) leave the quantization surface invariant.
Hence, as they only connect dynamical variables at one time (i.e.\, on the
surface), they don't contain any dynamics or
interaction dependent parts and the
interaction does not appear in the canonical commutation relations. The
boost generators are interaction dependent. Hence usually
approximations in the ordinary-coordinate-formalism will violate the
boost invariance.

On the light-cone the kinematical generators that leave the quantization
surface invariant are (different) shift operators ($P_-$, $P_1$, $P_2$),
rotations along the z-axis ($J_{12}$), longitudinal boost
($J_{+-}=J_{30}$) and two linear combinations of the remaining boosts and
rotations ($J_{-1}$, $J_{-2}$). Note that even the number of kinematical
generators has changed\footnote{In the light-cone formulation the number
of kinematical generators is maximum.} to 7. Here spatial rotations are
dependent on the dynamics. Hence often approximations on the
light-cone violate rotation invariance.

Note that the dynamics, i.e. evolution in the variable $x^+$, is
generated by $H=P^-=P_+$ rather than $P^0$ that would do the job in
ordinary coordinates.

\subsection{Simplicity of the light-cone vacuum}
It is generally believed that the true vacuum of QCD in ordinary
coordinates ($P^0 \vert vac > =0$) is very much
different from the perturbative vacuum state ($P^0_{free}\vert PTvac
>=0$). The complicated vacuum is
believed to give rise to confinement and spontaneous symmetry breaking.
In general the Hamiltonian (a normal ordered product of free field
creation and destruction operators\footnote{Note that by definition a
perturbative vacuum is anihilated by the destruction operators and hence
single free field operators has a vanishing perturbative-vacuum
expectation value.}) contains terms solely made out of
{\it creation} operators with the sum of their momenta being zero. So the
 the perturbative vacuum can not be an eigenstate of
the full Hamiltonian (i.e. the true vacuum being non-trivial) and
free fields must have positive and negative momenta. In
ordinary coordinates we are used to left and right movers.

On the other hand the light-cone dispersion relation doesn't allow for
positive energy ($P^-$) states with negative longitudinal momentum
($P^+$). Hence there are no terms in the Hamiltonian purely made out of
creation operators.
Hence on the light-cone the perturbative vacuum is always an eigenstate
of the full Hamiltonian (with zero eigenvalue).

Where have the broken phases and non-perturbative features gone? The
zero (momentum) modes are obviously not properly accounted for in the
above discussion. However they are not independent quantum degreees of freedom,
but rather are determined by a constraint equation in terms of the other
degrees of freedom. Now the non-trivial physics
resides in the constraint equation. This will be the topic of the next
section.

\section{Specific Example: $\phi^4$ in 1+1 Dimensions}
\subsection{General Theory} 
In this section we study the $\phi ^{4}$-theory in 1+1 dimensions 
as the simplest field theory which shows the mechanism of spontaneous 
symmetry breaking.

The ($\phi ^{4})_{1+1}$ theory is believed to have a second-order 
phase transition [6] for some critical coupling parameter $ 
\lc>0$. For $ \lambda >\lc$, $\phi $(x) has a nonvanishing vacuum 
expectation value and we are interested in understanding how this 
can be compatible with  the trivial light-cone vacuum structure. 
 We will see that the existence of a nontrivial constraint 
equation for the zero mode replaces the difficulties related 
to the vacuum polarisation. With certain approximations we will 
see how this constraint equation leads to a phase transition and 
an nonvanishing vacuum expectation value in a certain domain of 
the coupling.

\vspace{1ex}
\noindent In light-cone coordinates the bare ($\phi ^{4})_{1+1}$Lagrangian
is 
\begin{equation}
{\cal L}=\partial _+\phi \partial _-\phi -\frac{\mu ^{2}}{2} \phi 
^{2}-\frac{\lambda }{4!} \phi ^{4}
\end{equation} 
We put the system in a box (an interval) of lengh {\em d }in $ x^
-$ direction and impose periodic boundary conditions. 
\begin{equation}
\phi \left( x\right) =\frac{1}{\sqrt{d}} \sum _{n=-\infty }^{\infty 
}q_{n}\left( x^+\right) e^{ik_{n}^+x^-},\hspace{3ex}k_{n}^+=\frac{2
\pi n}{d} 
\end{equation} 
If we try to quantize canonically we get $ p_{n}=ik_{-n}q_{-n}$. 
So the momenta are connected to the coordinates and we have to use 
the Dirac-Bergmann prescription to quantize the theory. Since this is
thoroughly
discussed in literature [7], [8],  we will skip this 
topic here.

In particular, we get $ p_{0}=0$. So $ q_{0}$ has no canonical momentum 
and the equation of motion for $ q_{0}$ appears as a constraint 
equation: $ \frac{\partial L}{\partial q_{0}} =0$ (if it is applied 
on a physical state). 
As a result we get the creation and annihilation operators $ a_{n
}=\sqrt{4\pi |n|} q_{n}$, $ n\neq 0$ with $ a_{-n}=a_{n}^{\dagger 
}$ and the standard commutation relations $ [a_{n},a_{m}^{\dagger 
}]=\delta _{n,m}$. With the useful notation $|0|$:=1 we have 

\begin{equation}
\phi \left( x^+,x^-\right) =\sum _{n=-\infty }^{\infty }\frac{a_{
n}\left( x^+\right) }{\sqrt{4\pi |n|}} e^{ik_{n}^+x^-}.
\end{equation} 

For $ a_{0}=\sqrt{4\pi } q_{0}$ we get the constaint equation indicated 
above. It can be rewritten as $ \frac{\partial P^-}{\partial a_{
0}} =0$ where $ P^-$ is the light-cone Hamiltonian. This constraint 
equation can easily be motivated starting from the equation of motion 

\begin{equation}
\partial _+\partial _-\phi =-\mu ^{2}\phi -\frac{\lambda }{3!} \phi 
^{3}\hspace{2ex}.
\end{equation} 

Due to the periodic boundary conditions we get $ \int _{-\frac{d}
{2}}^{\frac{d}{2}}dx^-\partial _-(\partial _+\phi )=0$ [9]. Thus the 
constraint is the integral of the potential. 

\begin{equation}
0=\int _{-\frac{d}{2}}^{\frac{d}{2}}dx^-\mu ^{2}\phi +\frac{\lambda 
}{3!} \phi ^{3}.
\end{equation} 

If we seperate the zero mode $ \phi =\phi _{0}+\varphi $ we get 

\begin{equation}
0=\mu ^{2}\phi _{0}+\frac{\lambda }{3!d} \int _{-\frac{d}{2}}^{\frac{d
}{2}}dx^-\left( \phi _{0}+\varphi \right) ^{3}.
\end{equation} 

This is a complicated operator equation for $ a_{0}$ in terms of 
all the other modes in the theory. It refers to the complicated 
structure of the theory and due to this equation we can get a  
non-zero expectation value $ \langle 0|a_{0}|0\rangle \neq 0$ of $ \phi 
(x)$ in spite of a trivial vacuum. 

Spontaneous symmetry breaking 
is discribed in equal time quantisation by multiple vacua.
The choice of 
the vacuum defines the theory and the vacuum structure is complicated.
On the light-cone the vacuum is trivial but we get a complicated 
constraint equation for the zero mode $ a_{0}$. This equation has 
in general multiple solutions and the choice of a specific one defines 
the theory. 

\vspace{1ex}
\noindent In the next subsections we will derive the light-cone Hamiltonian
and look in more detail at the constraint equation. We will solve 
it approximately and show how, in principle, spontaneous symmetry 
breaking can arise. 

\subsection{The light-cone Hamiltonian} 
Now we can calculate the Hamiltonian $ P^-$ for the light-cone ($
\phi ^{4})_{1+1}$-theory. 

\begin{eqnarray}
P^-&=&\left( \frac{\lambda d}{96\pi ^{2}}\right) \left( \frac{g}{2
} \sum _{n=-\infty }^{\infty }\frac{a_{n}a_{-n}}{|n|} +\frac{1}{
4!} \sum _{k,\ell ,m,n}\frac{\delta _{k+\ell +m+n,0}}{\sqrt{|k\| l
\| m\| n|}} a_{k}a_{\ell }a_{m}a_{n}\right) \nonumber \\ 
&&\hbox{where }g=\frac{24\pi \mu ^{2}}{\lambda } ,\hspace{3ex}|0
|\hspace{.7ex}\raisebox{.065ex}{:}\hspace{-.5ex}=1
\end{eqnarray} 

Now we have to care for the divergences in the theory. In a two-dimensions
the only divergencies are tadpoles which can be removed by 
normal ordering. But it is not clear how to normal order the terms 
involving $ a_{0}$ since $ a_{0}$ itself has a complicated operator 
structure. For very general arguments [10] we use symmetric 
ordering for these terms. Since $ \langle 0|a_{0}|0\rangle $ may 
be different from zero, terms of the form $ a_{n}a_{0}^{2}a_{n}^{
\dagger }$, $ a_{0}a_{n}a_{0}a_{n}^{\dagger }$, \dots  may still 
give additional divergencies and require the further substraction 
\begin{equation}
\frac{1}{4}\sum _{n\neq 0}\frac{1}{|n|} \left( -6a_{0}^{2}\right) 
\end{equation} 
in Eq.\ (\ref{Pm}). 
If we are looking to the one mode approximation, as we will do in 
the next subsection, this Hamiltonian is finite. However, including 
all modes, some divergencies may still remain.

So $ P^-$ has the form 

\begin{eqnarray}
P^-&=&\left( \frac{\lambda d}{96\pi ^{2}}\right) \left( \frac{g}{2
} \sum _{n\neq 0}:\frac{a_{n}a_{-n}}{|n|} :+\frac{1}{4} \sum _{k
,\ell ,m,n}\frac{\delta _{k+\ell +m+n,0}}{\sqrt{|k\| l\| m\| n|}
} :a_{k}a_{\ell }a_{m}a_{n}:\nonumber \right.\\ 
\left.\right.&\left.\right.&\left.+\frac{1}{4} \sum _{n\neq 0} \frac{
1}{|n|} \left( a_{0}^{2}a_{n}a_{-n}+a_{n}a_{-n}a_{0}^{2}+a_{n}a_{
0}^{2}a_{-n}+a_{n}a_{0}a_{-n}a_{0}\nonumber \right.\right.\\ 
\left.\left.\right.\right.&\left.\left.\right.\right.&\left.\left.
+a_{0}a_{n}a_{0}a_{-n}+a_{0}a_{n}a_{-n}a_{0}-6a_{0}^{2}\right) +
\frac{g}{2}a_{0}^{2} +\frac{1}{4}a_{0}^{4} +\frac{1}{4}\nonumber 
\right.\\ 
\left.\right.&\left.\right.&\left.\cdot \sum _{k,\ell ,m\neq 0}\frac{
\delta _{k+\ell +m,0}}{\sqrt{|k\| l\| m|}} \left( a_{0}a_{k}a_{\ell 
}a_{m}+a_{k}a_{0}a_{\ell }a_{m}+a_{k}a_{\ell }a_{0}a_{m}+a_{k}a_{
\ell }a_{m}a_{0}\right) \right) \nonumber \\ 
\label{Pm}&&
\end{eqnarray} 
where we have separated out all the $ a_{0}$ terms.

\subsection{Solving the Constraint Equation [11]} 
The constraint equation is now given by $ \frac{\partial P^-}{\partial a
_{0}} =0$: 

\begin{eqnarray}
0&=&ga_{0}+a_{0}^{3}+\sum _{n\neq 0} \frac{1}{|n|} \left( a_{0}a_{
n}a_{-n}+a_{n}a_{-n}a_{0}+a_{n}a_{0}a_{-n}-3a_{0}\right) \nonumber 
\\ 
&&+\sum _{k,\ell ,m\neq 0} \frac{\delta _{k+\ell +m,0}}{\sqrt{|k\| l
\| m|}} a_{k}a_{\ell }a_{m}
\end{eqnarray} 

To initially solve this equation we have to make the approximations given
in 1. below, and we only look for solution of the type described in 2.

\begin{enumerate}\item Since the spontaneous symmetry breaking is 
a low energy effect it is reasonable to assume that the lowest energy 
mode will give the most important contribution to $ \langle 0|a_{
0}|0\rangle $. Therefore we will truncate the constraint equation 
to one mode $ a_{1}=a$. 
\item The higher lying levels should not be affected very strongly 
by the mechanism of spontaneous symmetry breaking. We only
look for a solution of this type.  Formally this type of solution satisfies
(16) below.

\end{enumerate} 

If we restrict the constraint equation to one mode we get: 

\begin{equation}
0=ga_{0}+a_{0}^{3}+2a_{0}a^{\dagger }a+2a^{\dagger }aa_{0}+a^{\dagger 
}a_{0}a+aa_{0}a^{\dagger }-a_{0}.
\end{equation} 

\noindent The solution will preserve particle number 
because of conserves momentum. We are particularly interested into the vacuum
expectation 
value 

\begin{equation}
\langle 0|\phi |0\rangle \hspace{2ex}=\hspace{2ex}\frac{\langle 0
|a_{0}|0\rangle }{\sqrt{4\pi }} \hspace{2ex}=\hspace{-.65ex}\raisebox
{.05ex}{:}\hspace{.7ex}\hspace{2ex}\frac{f_{0}}{\sqrt{4\pi }}
\hspace{1ex}
.
\end{equation} 

\noindent $ a_{0}$ must a function of $ N=a^{\dagger }a$ thus, 

\begin{equation}
a_{0}=\sum _{k=0}^{\infty }b_{k}N^{k}\hspace{1ex}.
\end{equation} 

\noindent This implies that $ a_{0}$ is diagonal and it can be written as

\begin{equation}
a_{0}=\sum _{k=0}^{\infty }f_{k}|k\rangle \langle k|\hspace{1ex},
\hspace{2ex}f_{k}\hspace{.7ex}\raisebox{.065ex}{:}\hspace{-.5ex}=
\langle k|a_{0}|k\rangle \hspace{1ex}.
\end{equation} 

\noindent Substituting this into the constraint equation and sandwiching it
between Fock states we get the following non linear finite difference 
equation: 

\begin{equation}
0=gf_{n}+f_{n}^{3}+\left( 4n-1\right) f_{n}+\left( n+1\right) f_{
n+1}+nf_{n-1}
\end{equation}

Since $ \langle N|a_{0}|N\rangle =0$ for $ \lambda =0$ and because 
of the condition 2 mentioned above we get $ \langle N|a_{0}|N\rangle 
\rightarrow 0$ for $ N\rightarrow \infty $ and all $\lambda $. So 
we search for a solution with the property  

\begin{equation}
\label{lim} \lim_{n\rightarrow \infty } f_{n}=0
\end{equation} 

\noindent Therefore we study the large {\em n }behavior of our equation and
drop the $ f_{n}^{3}$ term: 
\begin{equation}
f_{n+1}+4f_{n}+f_{n-1}=0
\end{equation} 
which has the asymptotic behavior $ f_{n}\propto c^{n}$, with $ c
^{2}+4c+1=0$ or $ c=-2\pm \sqrt{3}$.
Because of (\ref{lim}) we have to reject the $ (-2-\sqrt{3} )^{n}
$ solution. As we will see shortly this is only possible for $ g
\le \gc $ (except for the trivial solution 
$ f_{n}\equiv 0$, which does always exist).

To calculate the critical point we start from a small $ f_{0}$, 
since we are looking for solutions close to the trivial one. So 
we still can drop the $ f_{n}^{3}$ term. We are left with the linear 
equation ($ f_{-1}=0=f_{-2}$) 
\begin{equation}
\left( 4n+g-1\right) f_{n}+\left( n+1\right) f_{n+1}+nf_{n-1}=0
\end{equation} 
To solve this equation we introduce the generating function  
\begin{equation}
F\left( z\right) =\sum _{n=0}^{\infty }f_{n}z^{n}\hspace{1ex}.
\end{equation} 
Our difference equation for $ f_{n}$ gives us a differential equation 
for $ F(z)$: 
\begin{equation}
\left( z^{2}+4z+1\right) F'\left( z\right) +\left( z+g-1\right) F
\left( z\right) =0
\end{equation} 
This equation can easily be solved and the solution is 
\begin{equation}
\label{andi} F\left( z\right) =F\left( 0\right) \left( \frac{z+2-
\sqrt{3}}{2-\sqrt{3}}\right) ^{-\frac{\sqrt{3} -3+g}{2\sqrt{3}} 
}\left( \frac{z+2+\sqrt{3}}{2+\sqrt{3}}\right) ^{-\frac{\sqrt{3} 
+3-g}{2\sqrt{3}} }
\end{equation} 
If $ f_{n}$ goes asymptotically like $ (-2+\sqrt{3} )^{n}$ then the 
radius of convergence for $ F(z)$ is $ R=\frac{1}{|-2+\sqrt{3} |
}=2+\sqrt{3}$ else it is $ R=\frac{1}{|-2-\sqrt{3} |}=2-\sqrt{3}
$.

So we can reformulate our condition $ f_{n}\rightarrow 0$, $ n\rightarrow
\infty $ as follows: $ F(z)$ has no singularity in the unit disc 
of the complex plane.
{}From (\ref{andi}) we read off that $ F(z)$ has a singularity at $ z
=-2+\sqrt{3}$ unless $ -\frac{\sqrt{3} -3+g}{2\sqrt{3}}$ is a non-negative
integer. 
\begin{eqnarray}
&&-\frac{\sqrt{3} -3+g}{2\sqrt{3}} =k\hspace{1ex},\hspace{2ex}k=0
,1,2,3,\ldots \nonumber \\ 
&\Leftrightarrow &g\hspace{1ex}=\hspace{1ex}3-\sqrt{3} -2\sqrt{3} 
k
\end{eqnarray} 
The only solution with positive {\em g }is 
\begin{equation}
\gc=3-\sqrt{3} =\frac{24\pi \mu ^{2}}{\lc} 
\end{equation} 
or 
\begin{equation}
\frac{\lc }{\mu ^{2}} =4\pi \left( 3+\sqrt{3}\right) \approx 59.5
\hspace{1ex}.
\end{equation} 
This value agrees well with the numerical result
for the equal time theory  [12]
\begin{equation}
\frac{\lc}{\mu ^{2}} \approx 30 - 60\hspace{1ex}.
\end{equation} 
 This agreement should nevertheless not be overestimated since the 
numeric value for $ \frac{\lc}{\mu ^{2}}$ does not take
higher modes into account. Moreover the calculations 
with higher modes seem to diverge logarithmically which reflects 
the fact that the renormalisation is still missing.
A more complete discussion including higher modes is given in [13]

To explore whether or not $ \gc$ is an isolated point or the beginning 
of a continuum of critical couplings we have to look at the full 
constraint equation including the $ f_{n}^{3}$ term. 

\subsubsection{The $\delta $-Expansion} 
A powerful analytical method to linearize non-linear difference equations
is the $\delta $-expansion. We rewrite 
\begin{eqnarray}
f_{n}^{3}&=&f_{n}^{1+2\delta }\approx f_{n}\left( 1+\delta \ln f_{
n}^{2}\right) \hspace{1ex},\nonumber \\ 
g&=&g^{\left( 0\right) }+\delta g^{\left( 1\right) }+\ldots \hspace{1ex}
,\hspace{2ex}f_{n}=f_{n}^{\left( 0\right) }+\delta f_{n}^{\left( 1
\right) }+\ldots 
\end{eqnarray} 

\noindent and solve the difference equation as an expansion in $\delta $ [11].
The result for the first order and $ \delta =1$ is 
\begin{equation}
\gc =\left( 2-\sqrt{3}\right) \left( 1+\frac{1}{\sqrt{3}} \ln \left( 2
+\sqrt{3}\right) \right) -\ln f_{0}^{2}\hspace{1ex}.
\end{equation} 
This is quite a good approximation to the exact result for $ g<1$. 
It is shown as the dashed line in Fig.\ 1.

\subsubsection{Numerical Solution} 
The difference equation can easily be calculated. To 
find the stable solution $ (f_{n}\rightarrow 0)$ we truncate the 
series by setting $ f_{N+1}=0$ for some large $ N(\approx 10)$. We get 
a system of equations 
\begin{eqnarray}
0&=&\left( \gc -1\right) f_{0}+f_{0}^{3}+f_{1}\nonumber \\ 
0&=&\left( \gc +3\right) f_{1}+f_{1}^{3}+2f_{2}+f_{0}\nonumber \\
&&\hspace{2ex}\vdots \nonumber \\ 
\label{ooo} 0&=&\left( \gc +4N-1\right) f_N+f_N^{3}+Nf_{N-1}
\end{eqnarray} 
and solve it numerically for $ f_{0}$ as a function of $ g_{\rm critical}
$. The result is indicated by the solid line in Figs.\ 1 and 2. 
At $ \lambda =\lc $ a second order phase transition can be seen. 
The domain of negative {\em g }may be of interest in the Higgs model 
which starts from a negative mass $\mu ^{2}$. The structure of solutions 
seems to be more involved in this domain, Fig 2.

\subsection{Calculation of the Eigenvalues of $ P^-$} 
We can also study the eigenvalues of the light-cone Hamiltonian $ P^
-$ truncated to the one mode problem.

Since $ P^-$ conserves momentum it is diagonal in the number operator 
{\em N }so that the energy eigenstates are eigenstates of {\em N}. 
\begin{eqnarray}
E_{n}=\langle n|H|n\rangle &=&\frac{3}{2}n\left( n-1\right) +ng-\frac{f
_{n}^{4}}{4}-\frac{2n+1}{4}f_{n}^{2}\nonumber \\ 
&&+\frac{n+1}{4}f_{n+1}^{2} +\frac{n}{4}f_{n-1}^{2}
\end{eqnarray}

The result for the first three eigenvalues is shown in Fig.\ 3 by 
the solid curve. The dashed lines show the result for $ a_{0}=0$. 
For $ g>g_{\rm critical}$ the two lines coincide but at $ g=g_{\rm
critical}
$ there is a phase transition and the energy of the first excited state 
decreases as {\em g }is decreased. For large {\em N }this effect is very 
small as we would expect from our approximations.  We note that the mass gap is
a minimum at the critical point. 

\section{Work in Progress} 
The next step is to take several modes into account and this is discussed
in detail in reference [13].There is still an outstanding problem of operator
ordering and renormalisation
that has to be solved to get the complete solution 

\section{Conclusions} 
\begin{itemize}\item light-cone field theories may show the feature 
of spontaneous symmetry breaking inspite of their trivial vacuum 
structure. This effect results from the complicated properties of 
the zero modes in the theory. The zero mode is connected to all 
the other modes in the theory by the constraint equation. 
\item  For $ 1+1$-dimensional $\phi ^{4}$-theory the solution of 
the one mode approximation gives the explicite result $ \lc =4\pi 
(3+\sqrt{3} )\mu ^{2}$ which lies close to the numerical result 
for equal time theories. At this point a second 
order phase transition takes place.  
\item For $ \lambda <\lc $ the theory has the same spectrum as it 
would have without the zero mode. For $ \lambda >\lc $ the energy 
of the first excited state is poositive with a minimium at $ \gc $
The $ E_{n}$ for large {\em n 
}are nearly unaffected. 
\item In general $ a_{0}$ gives rise to an infinite number of new 
interactions in the effective Hamiltonian even if $ \lambda <\lc 
$. These interactions are determined by the constraint equation.
See [13] for more details.
\end{itemize} 

\section{References}

\begin{itemize}

\item[1.] S. J. Brodsky, G. McCartor, H. C. Pauli and S. S. Pinsky, Particle
World {\bf 3} (1992) 109;  X. Ji, Comments Nucl.\ Phys.\ {\bf 21} (1993) 123.

\item[2.] S.Fubini and G.Furlan, Physics {\bf 1} (1965) 229;
                  S.Weinberg, Phys. Rev. {\bf 150} (1966) 1313.

\item[3.] P.Dirac, Rev. Mod. Phys. {\bf 21} (1949) 392.

\item[4.] H.Leutweyler and L.Stern, Ann. Phys. {\bf 112} (1978) 94.

\item[5.] S.Chang, R.Root and T.Yan, Phys. Rev. {\bf D7} (1973) 1133.

\item [6.] B. Simon, R. B. Griffiths, Commun.\ Math.\ Phys.\ {\bf 33}, 145
(1973).

\item [7.]T. Maskawa, K. Yamawaki; Prog. Theor. Phys. {\bf 56} (1976) 270;
R.S. Wittman, in Nuclear and Particle Physics on the light-cone, M. B. Johnson
and L. S. Kisslinger, eds. (World Scientific).

\item [8.] T. Heinzl, S. Krusche, S. Simberger, E. Werner,
{\sl Nonperturbative light-cone Quantum Field Theory Beyond The
 Tree
Level}, Regensburg preprint TPR 92-16; Heinzl,
S. Krusche, and E. Werner, Phys. Lett, {\bf B256},
55 (1991); TPR 91-23 (1991).

\item [9.] D. Robertson Phys. Rev. {\bf D47} 2549 (1993).
G. McCartor and D. G. Robertson Z. Phys.
{\bf C53} 53 (1992) 679.

\item [10.] C. M. Bender, L. R. Mead, S. S. Pinsky, Phys.\ Rev.\ Lett.\
{\bf 56} (1986) 2445.

\item [11.]  C. M. Bender, S. S. Pinsky, B. van de Sande,
Phys. Rev. {\bf D48} 816 (1993).

\item [12.] S. J. Chang, Phys.\ Rev.\ D {\bf 13}, 2778 (1976).M. Funke, V.
Kavlfass,
and H. Kummel, Phys. Rev. {\bf D35} 35, 621 (1987);
H. Kroger, R. Girard
and B. Dufour, Phys. Rev. {\bf D35L}(1987) 3944.

\item [13.] S. Pinsky; B. van de Sande OHSTPY-HEP-93-(  ) Spontaneous Symmetry
Breaking of (1 + 1) Dimensional $\phi^4$ theory in Light-Front Field Theory II.

\item[14.] H. C. Pauli MPIH-V-1991; A. C. Kalloniatie and H. C. Pauli
MPIH-V6-93

\end{itemize}

\section{Figure Captions}

\begin{description}
\item[Figure 1.] $g=24\pi\mu^2/\lambda$ vs.\ $f_0=\sqrt{4\pi}\langle 0|\phi
(x)|0\rangle$. The solid curve obtained from the numerical solution of
(\ref{ooo}) with $N=10$. The dashed curve is the critical curve obtained from
the first-order $\delta$-expansion.
\item[Figure 2.] The critical curve obtained numerically shows an interesting
behavior in the negative $g$ domain.
\item[Figure 3.] The lowest three energy eigenvalues as a function of $g$ from
the
numerical solution of (\ref{ooo}) with $N=10$. The dashed line is the symmetric
solution $f_0=0$ and the solid line is the solution with $f_0\neq 0$ for
$g<g_{\rm
critical}$.

\end{description}
\end{document}